# Analysis of the attack and its solution in wireless sensor networks

Ali Taghavirashidizadeh* , Arash Bahram Zarei , Arman Farsi

1. Department of Electrical and Electronics Engineering, Islamic Azad University Central Tehran Branch(IAUCTB), Tehran, Iran.
2. Undergraduate Student of Software Engineering, Shahid Yazdan Panah University of Sanandaj.
3. Master of science biomedical engineering, Marche Polytechnic University.

* Email: ali.taghavi.eng@iauctb.ac.ir

## Abstract

Several years ago, wireless sensor networks were used only by the military. These networks, which have many uses and are subject to limitations, the most important of which is the energy constraint, this energy constraint creates the requirement that the number And the length of the messages exchanged between the sensors is low. Sensor networks do not have a stable topology due to their inaccessible environments, and continually changes with the disappearance or addition of a node, support for the topology is performed in three steps before deployment, after deployment and deployment. . The sensor nodes are scattered across the field and transmitted by a multidimensional connection to the sink. The communication protocol of the sensor networks used by all nodes in the network and sink. The protocol consists of five layers and three levels of management. Sensor networks require a kind of security mechanism due to inaccessibility and protection. Conventional security mechanisms are inefficient due to the inherent limitations of sensor nodes in these networks. Sensor nodes, due to energy and resource constraints, require security requirements such as the confidentiality of data integrity data, authentication, synchronization, etc. Currently, many organizations use wireless sensor networks for purposes such as air, Pollution, Traffic Control and Healthcare Security is the main concern of wireless sensor networks. In this article, I will focus on the types of security attacks and their detection. This article outlines security needs and security attacks in wireless sensor networks. Also, the security criteria in wireless sensor networks are mentioned.

## 1. Introduction

Wireless sensor network security is a subject discussed several years ago. Networks have different applications. These programs are monitored, tracked and controlled in several steps. Wireless sensor networks include a large number of small nodes. These nodes are located in some of the important areas. There is a group of applications that are used for some purposes. Therefore, in military applications, sensor nodes include monitoring, battlefield monitoring and object tracking. Medical applications, sensors can be useful in patient diagnosis and monitoring. Most These programs are deployed to monitor a region and, if captured by a critical agent, respond wireless sensor networks as a new central stage in the IT ecosystem and a rich active research area including hardware and design design , Distributed algorithms, programming models, data management, security and reactive reactions A record of popular will appear. Wireless sensor networks are widely used in a region to be explored for a specific task. Sensing nodes are responsible for physical recording. The main purpose of this paper is to find a security solution for energy efficiency to enable sensor networks Protecting Wireless from Any Attacks Also, this article offers several resolutions for wireless sensor networks.

## 2. Wireless Sensor Network

A wireless sensor network is referred to as a self-directed sensor that is spaced apart and used to measure a node of some physical quantities with environmental conditions such as temperature, sound, vibration, movement pressure, or pollutants in different pollutant locations. . In addition to one or more sensors, each network node is usually equipped with a radio transmitter and receiver (or any other wireless communication device), a small microcontroller, and an energy source (usually a battery). The size of a sensor node varies depending on the size of its packaging, and it can be narrowed down to single gravel; the pieces of the microscope must still be made. Similarly, the price of each sensor node can range from a few hundred dollars to a few cents, depending on the size and complexity of a node. Limitations in price and size in sensor nodes lead to constraints on sources such as energy, memory, processing speed and bandwidth. One of the important requirements of the sensor network is the concurrent service. The importance of time in sensor networks has led to the disruption of sensor synchronization as one of the primary goals of the enemy to attack these networks. The enemy tries to prevent proper synchronization in the network in various ways, such as disrupting synchronization messages, changing or forging them, delaying time-sensitive messages, conquering some nodes, and sending misleading messages by them. Despite the introduction of several synchronization methods for sensor networks in recent years, so far there is no comprehensive synchronization method that can meet the security and efficiency requirements of these networks simultaneously.

## 3. Introducing the overall structure of wireless sensor networks

In the design of wireless sensor sensors, sensor nodes have been used. Despite the various uses that exist for sensor nodes in wireless sensor networks, the main task of sensor nodes is sensing, collecting data and transmitting it to the sink. Sensor nodes Originals are the nodes that have the desired data. Due to the inherent constraints of the wireless sensor networks or the position of the sensor nodes relative to the sink, they can not communicate directly with a sink jump, thus requiring multiple paths. Fig. 1 shows the structure of this process. Sensor nodes are located. In addition to fulfilling the tasks, the source and sink nodes must, when necessary, be used as interfaces for the transmission of other sensing nodes. In order to achieve this goal, due to the constraints on the amount of resources (especially low-power and often non-rechargeable or replaceable) in wireless sensor networks, it requires optimized routing algorithms to provide high-performance storage. Energy can provide the best path from source to sink, in other words, it should maximize the lifetime of the entire network as much as possible [1].

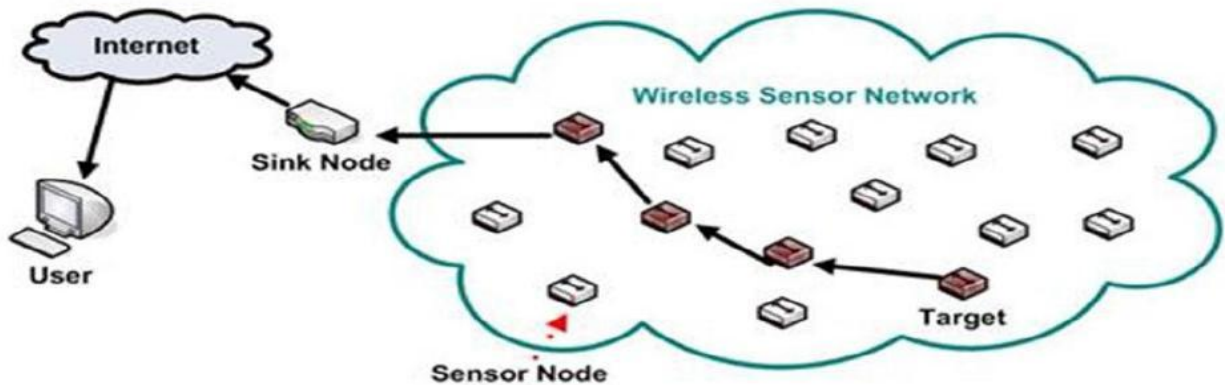

Figure 1: General structure of wireless sensor network [1]

## 4. Security requirements

Security in wireless sensor networks is a must-have requirement. These requirements not only guarantee the protection of sensor information, but also to obtain limited resources at each sensor node, where the sensor network is still alive. The motive, the attacker, and the attacker's opportunities are two factors that make it possible for an attacker to attack wireless sensor networks.

## 1-4- Confidentiality of information

The confidentiality of information keeps secret information from enemies. The perfect way to keep the invisible information is to encrypt the data with a secret key.

## 2-4- Data authentication

The basis of multiple applications in the wireless sensor network is the authentication of the message. Data authentication will disable any unauthorized network in the network and reveal the nodes that are subject to the unauthorized condition. It is also important that the data is started with the assurance of a correct source and the correct node must be the end of the connection.

## 3-4- Data integrity

The enemy nodes attempt to change or substitute, so data integrity ensures that the recipient of the message received by the unauthorized transfer has not changed.

## 4-4- Fresh information

New data means that the information is new. This ensures that no old data or messages are broadcast. Although the confidentiality of the data is assured, it is necessary to ensure that novelty is achieved for each message, and that the repeated message is not broadcast, in addition to the malicious node does not send duplicate data.

## 5-4- Access control

Access control prevents unauthorized access to a source. This action should be able to prevent unauthorized access to the network.

## 5. Security attacks in wireless sensor networks

Wireless sensor networks are very weak and sensitive to various types of security attacks that trigger the broadcast. Another reason why the sensor nodes are dangerous is that they are placed in dangerous environments like the battlefield. Security threats and attacks on wireless sensor networks are as follows:

## 1-5- Sybil attack:

Wireless sensor network is vulnerable to Sybil attack. In such cases, the node can have more than one identity using other nodes, so a node only provides several identities to network nodes. Figure 2. The sybil attack tries to eliminate the data integrity, security, and resource utilization that the distributed algorithm tries to achieve. Authentication and encryption mechanisms can prevent a remote host and Sybil attack to a wireless sensor network. Public key cryptography can prevent such personal attacks, but it also costs a lot of resources in sensor networks. The identity of the sensor node should be confirmed, so Carloff and Wagner say that it can be done using public key authentication encryption, but creating and verifying digital signatures is beyond the capabilities of sensor nodes [2,3].

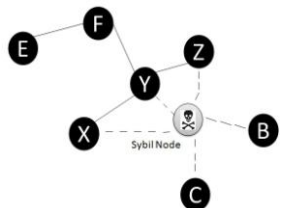

**Figure 2: Sybil attack**

## 2-5- Wormhole attack

A wormhole attack is one of the most significant and most dangerous attacks on the wireless sensor network. In this type of attack, the attacker's node connects two points of the network with a relatively fast communication link, called wormhole. Then, through this tunnel, the network sends the network from one network point to its neighbor at another point. In the Wormhole attack, the attacker uses tunes between himself and other nodes to confuse the routing protocol. The attacker captures the packets in a location from the network and takes them to another location through a tunnel. Tunneling or retransmission of bits can be done selectively. Using the techniques of signal processing, you can prevent a Wormhole attack:
If the data bits are modulated in a particular way, only the neighboring nodes are known. The use of directional antennas by mobile nodes can improve security. Neighboring nodes examine the signals received from each other and share common evidence when neighbors are verified. Fig. 3.

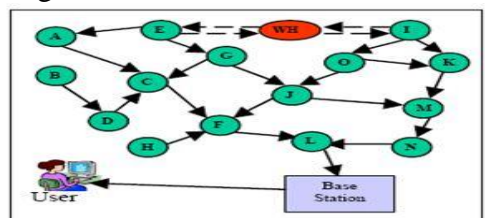

**Figure 3: Attack wormhole [2]**

## 3-5- HELLO Flooding attack

In this attack, which is one of the easiest attacks on the wireless sensor network, the attacker sends the Hello packet to the highest transmission power for the transmitter or receiver. The message receiving nodes assume that the sender node is close to them and Sends packets to nodes. This attack on the network causes an increase and a specific node of the denial of service attack. Figure 4.

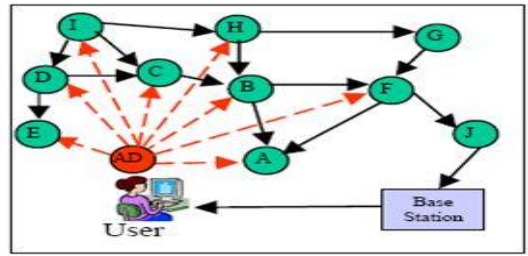

**Figure 4: Attack Hello Flood [3]**

## 4-5- Sinkhole attack

In this type of attack, the attacker shows himself in the grid as a high resource resource node, which always has the shortest route with this attribute, and thus all the data passes through the attacker's node [4,5]. Figure 5.

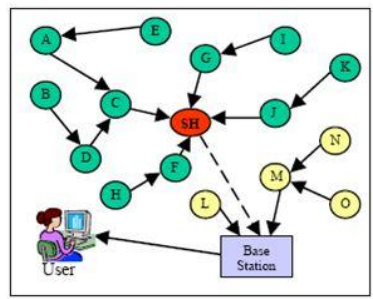

**Figure 5: Sinkhole Attack**

## 6. Security Assessments for Wireless Sensor Networks

In recent years, wireless sensor networks have grown vastly and are being used for a wide range of applications such as climate, military targeting and patient monitoring. Therefore, these sensor networks need to protect the illegal attackers. There are some security measures that sensor networks need to have [5].

## 1-6- Encryption

In fact, most wireless sensor networks are located in an open area or risk location and are thus sensitive to network attacks. Eavesdropping or adding messages to the network is significant for wireless sensor networks. Contrary to a word Normally, the key does not allow the user to directly access the information. Instead, a key encryption algorithm can initially encrypt encrypted data. In fact, without encryption, encrypted data is unavailable [6,5].

## 2-6- Division of information

The partitioning technique is to divide the data into networks in several or more sections, dividing the data into multiple packets so that each packet is transferred to other nodes in a different direction. At this point, the attacker tries to receive all data packets from the network, so it must be accessible to all networks. This is a complete solution, and energy consumption is still normal [7,5].

## 3-6- Safe data collection

The data transmitted in the wireless sensor network has increased compared to the previous one. As a result, the biggest problem in the network is traffic. So the cost increases. To reduce the cost and network traffic, the wireless sensor node performs measurements before transferring to the base station. The wireless sensor network architecture is such that data collection is done in many places on the network and that the collection sites Information should be secure [7].

## 4-6- Cryptography

One of the most important issues in network security and cryptographic computing is the knowledge encryption that examines the principles and methods of transferring or storing information securely, even if the data transmission paths and communication channels or insecure information stores are unsafe. Information and privacy protection This means that only the sender and receiver can understand the content of the message, others may be able to see the content, but in their view their content should be completely obscure. [8]

## 7. Conclusion

The security issue in wireless sensor network is more important than other issues. In recent years, security in wireless sensor networks has been significant. Wireless sensor networks are growing and used in commercial, health and military environments. In the security and privacy debate, unfortunately, there are still ways to influence these networks. Although the security of this area is now upgraded and network penetration has become more difficult, a successful attack is enough to question the credibility of a system, hence encryption is one of the most important issues in network and computer security. This field requires more cryptographic algorithms In this paper, we tried to examine the different angles of wireless sensor networks and attacks on wireless sensor networks, in particular with regard to security and encryption, it was attempted to include a large part of this research.

# References

[1] Abhishek Jain, Kamal Kant and M. R. Tripathy, "Security Solutions for Wireless Sensor Networks", to appear in IEEE ICACCT 2012.

[2] Al-Sakib Khan Pathan, Hyung-Woo Lee, Choong Sean Hong, "Security in Wireless Sensor Networks: Issues and Challenges", Proc. ICACT 2006, Volume 1, 20-22 Feb, 2006, pp. 1043-1048.

[3] Culler, D. E and Hong, W., "Wireless Sensor Networks", Communication of the ACM, Val. 47, No. 6, June 2004, pp. 30-33.

[4] Kalpana Sharma. M K Ghose, "Wireless Sensor Networks: An Overview on its Security Threats", IJCA Special Issue on Mobile Adhoc Networks 2010.

[5] Yan-Xiao Li, Lian-Qin and Qian-Liang, "Research on Wireless Sensor Network Security", In Proceedings of the International Conference on Computing and Security, 2010 IEEE.

[6] V. Thiruppathy Kesavan and S. Radhakrishnan, "Secret Key Cryptography Based Security Approach for Wireless Sensor Networks", International Conference on Recent Advances in Computing and Software Systems, 2012 IEEE.

[7] Kalpana Sharma. M K Ghose, Deepak Kumar, Raja Peeyush Kumar Singh, Vikas Kumar Pandey, "A comparative Study of Various Security Approaches Used in Wireless Sensor Networks", In IJAST, Vol 7, April 2010.

[8] Pathan, A. S. K., Hyung-Woo Lee, and Choong Seon Hong "Security in Wireless Sensor Networks: Issues and Challenges" Advanced Communication Technology (ICACT), 2006.

[9] Mohsen Ahmadi, Ali Taghavirashidizadeh, Danial Javaheri, Armin Masoumian, Saeid Jafarzadeh Ghoushchi, Yaghoub Pourasad "DQRE-SCnet: a novel hybrid approach for selecting users in federated learning with deep-Q-reinforcement learning based on spectral clustering", Journal of King Saud University-Computer and Information Sciences, Elsevier, 2021/8/27

[10] A Taghavirashidizadeh, R Parsibenehkohal, M Hayerikhiyavi, M Zahedi "A Genetic algorithm for multi-objective reconfiguration of balanced and unbalanced distribution systems in fuzzy framework", Journal of Critical Reviews, 2020

[11] Ali Gharamohammadi, Mohammad Reza Yousefi Darestani, Ali Taghavirashidizadeh, Ahmad Abbasi, Arash Shokouhmand "Electromagnetic sensor to detect objects by classification and neural networks", Sensor Letters, 2019/9/1

[12] Mahmoud Makkiabadi, Siamak Hoseinzadeh, Ali Taghavirashidizadeh, Mohsen Soleimaninezhad, Mohammadmahdi Kamyabi, Hassan Hajabdollahi, Meysam Majidi Nezhad, Giuseppe Piras "Performance Evaluation of Solar Power Plants: A Review and a Case Study", Processes, 2021/12/14

[13] Ali Taghavirashidizadeh, Fatemeh Sharifi, Seyed Amir Vahabi, Aslan Hejazi, Mehrnaz SaghabTorbati, Amin Salih Mohammed "WTD-PSD: Presentation of Novel Feature Extraction Method Based on Discrete Wavelet Transformation and Time-Dependent Power Spectrum Descriptors for Diagnosis of Alzheimer's Disease", Computational Intelligence and Neuroscience, 2022/5/11

[14] Mahmoudreza Moghimhanjani, Ali Taghavirashidizadeh "Analysis of liver cancer detection based on image processing", 7th International Conference On Modern Finding In Sciences And Technology With A Focused On Science In Development Services, 2020/7/21

[15] Nahid Ebrahimi, Ali Taghavirashidizadeh, Seyyed Saeed Hosseini "Extend the lifetime of wireless sensor networks by modifying cluster-based data collection", 3rd International Congress On Science And Engineering, 2020/3/14